\documentclass[11pt,a4paper]{article}

\usepackage{amsmath,amssymb}

\def\bee{\begin{eqnarray}}
\def\ene{\end{eqnarray}}

\begin{document}

\title{\bf Neutrino heating of a shock wave within magnetorotational model 
}

\author{A.~A.~Gvozdev$^a$\footnote{{\bf e-mail}: gvozdev@yniyar.ac.ru},
I.~S.~Ognev$^{a}$\footnote{{\bf e-mail}: ognev@yniyar.ac.ru}
\\
$^a$ \small{\em Yaroslavl State University } \\
\small{\em ul. Sovetskaya 14, Yaroslavl, 150000 Russia}
}

\date{}
\maketitle

\begin{abstract}
Based on the magnetorotational model of a supernova explosion with core 
collapse, we investigate the significant processes of neutrino heating of 
the supernova shock. These processes should be taken into account in 
self-consistent modeling, since the neutrino heating mechanism is capable 
of increasing the explosion efficiency. We show that, even in the presence 
of a strong magnetic field ($ B \sim 10^{15} $ G) in the shock formation 
region, the heating rate is determined with good accuracy by the 
absorption and emission of neutrinos in direct URCA processes. Moreover, 
the influence on them of a magnetic field is reduced to insignificant 
corrections. 
\end{abstract}

Explosions of supernovae with core collapse are known to be generally 
accompanied by an intense outward ejection of part of the material. 
However, an efficient explosion do not occur in the framework of the 
currently existing models.
Thus, for example, in the standard spherically symmetric 
supernova explosion model, the shock stops on a scale on the order of a 
hundred kilometers from the center of the remnant. Attempts to improve 
this model by applying relativistic corrections and using a 
self-consistent description of neutrino propagation (based on the solution 
of the Boltzmann equation) do not lead to a significant modification of 
the explosion pattern (Liebendoerfer et al. 2001). The currently existing 
2D calculations including the additional shock heating through convection 
and interaction with the neutrino flux do not lead to a successful 
supernova explosion either (Buras et al. 2003). On the other hand, the 
currently available observational data on several supernovae suggest that 
their explosions are asymmetric (Wheeler et al. 2002; Wang et al. 2001); 
moreover, this asymmetry can be relatively large (Leonard et al. 2000). It 
would be natural to assume that this asymmetry is the result of the rapid 
rotation of the collapse remnant or the presence of a strong magnetic 
field. Note that, according to existing models, the generation of a 
magnetic field in the remnant is directly related to its rapid rotation.

At present, the best-known supernova explosion model with a 
self-consistent allowance for the magnetic field is the so-called 
magnetorotational model by Bisnovatyi-Kogan (1970). The presence of a 
primary magnetic field and an angular velocity gradient in this model 
leads to the linear growth of a secondary magnetic field with time to a 
certain critical value. Once the latter has been reached, an axially 
symmetric (relative to the equatorial plane) supernova explosion occurs. 
However, as recent calculations by Ardeljan et al. (2004) showed, the 
linear growth of the magnetic field is disrupted by the development of 
magnetorotational instability. The development of this instability leads 
to a rapid growth of magnetic field perturbations to strengths 
$ B \sim 10^{15} - 10^{16} $ G, and to the formation of a shock.

As the magnetorotational instability develops, the kinetic energy of the 
rotation of the envelope with an angular velocity gradient transforms into 
the kinetic energy of the outward ejection of material through the rapidly 
growing magnetic field perturbations (Balbus and Hawley 1991, 1998). 
However, another additional energy source, shock heating by a neutrino 
flux, has long been known (Bethe and Wilson 1985). In the model under 
consideration, the neutrino heating mechanism is capable of increasing the 
explosion efficiency and is of particular interest.

The direct URCA processes 
\bee
\nu_e + n \to p + e^-  ,
\label{Ur1}
\\
\tilde\nu_e + p \to n + e^+  .
\label{Ur2}
\ene
are generally believed to be the dominant neutrino shock heating 
reactions. Another popular neutrino-lepton process, 
\bee
&& \nu_i + \tilde\nu_i \to e^+ + e^-  ,    
\label{Nu}
\\
&& (i = e, \mu, \tau)  ,
\nonumber
\ene
is inefficient far from the center, because the angle between the neutrino 
and antineutrino momenta is small. Note that in a medium with a strong 
magnetic field, the production processes of an $ e^+ e^- $ pair by a 
single neutrino,  
\bee
&& \nu_i \to\ \nu_i + e^+ + e^-  ,
\label{Nu1}
\\
&& \tilde\nu_i \to \tilde\nu_i + e^+ + e^-  ,   
\label{Nu2}
\\
&& (i = e, \mu, \tau) .
\nonumber
\ene
open up kinematically and can be important. In this paper, we compare the 
neutrino shock heating efficiencies in the presence of a strong magnetic 
field in the standard direct URCA processes and reactions 
(\ref{Nu1})-(\ref{Nu2}).

The neutrino heating rate per nucleon in the direct URCA processes 
(\ref{Ur1})-(\ref{Ur2}) can be calculated as  
\bee
Q_0^{\nu, \tilde\nu}
= 
\frac{1}{N_N} \int
\omega K_{\nu, \tilde\nu} f_{\nu, \tilde\nu} (\omega, \vec r) 
\frac{d^3k}{(2 \pi)^3}  , 
\label{Q0}
\ene
where $ k_\alpha = (\omega, \vec k) $ is 4-momentum of the (anti)neutrinos, 
$ f_{\nu, \tilde\nu} (\omega, \vec r) $ is their local distribution 
function, $N_N$ is the local nucleon number density, and  
$ K_{\nu, \tilde\nu} $ is the absorption coefficient defined as the rate 
of reactions (\ref{Ur1})-(\ref{Ur2}) underintegrated over the neutrinos. 
In what follows, we use the natural system of units with 
$ c = \hbar = k = 1 $.

In the case of a moderate magnetic field where the $ e^+ e^- $ plasma 
occupies many Landau levels 
($ \left< \omega^2_{\nu_e} \right> \gtrsim 2 eB $), its influence on the 
direct URCA processes is rather weak. For the absorption coefficient, we 
can in this case use its field-free expression. Assuming that the 
$ e^+ e^- $ plasma is ultrarelativistic and that the nonrelativistic nucleons 
have a Boltzmann distribution, we can represent the absorption coefficient 
as 
\bee
K_{\nu, \tilde\nu} 
=
\frac{G^2}{\pi}
(1 + 3 g_a^2) Y_{n, p} N_N \, 
\frac {\omega^2} { 1 + \exp[ (-\omega \pm \mu_e) / T ] }  .
\ene
Here, $ G = G_F \cos\theta_c $, where $G_F$ is the Fermi constant, 
$\theta_c$ is the Cabibbo angle,
$ g_a \simeq 1.26 $ is the axial constant of the charged nucleon current, 
$ Y_n = N_n / N_N $, $ Y_p = 1 - Y_n $, 
$ N_N = N_n + N_p $, $ \Delta = m_n - m_p $, 
where $ N_n, N_p, m_n, m_p $ are the local neutron and proton number 
densities and masses, respectively, 
$\mu_e$ is the chemical potential of the electrons, 
and $T$ is the local temperature.

It is convenient to represent formula (\ref{Q0}) for the heating rate of the 
medium in the direct URCA processes in terms of the mean quantities of 
neutrino radiation:  
\bee
&& \left< \omega_{\nu_e}^n \right> 
= 
\left( \int \omega^{n+1} f_{\nu_e} (\omega, \vec r) \frac{d^3k}{(2 \pi)^3} 
\right)
\left( \int \omega f_{\nu_e} (\omega, \vec r) \frac{d^3k}{(2 \pi)^3} 
\right)^{-1}   ,
\\
&& \left< \chi_{\nu_e} \right> 
= 
\left( \int \chi \omega f_{\nu_e} (\omega, \vec r) \frac{d^3k}{(2 \pi)^3} 
\right)
\left( \int \omega f_{\nu_e} (\omega, \vec r) \frac{d^3k}{(2 \pi)^3} 
\right)^{-1}   ,
\ene
(where $\chi$ is the cosine of the angle between the neutrino momentum and 
the radial direction) and the total neutrino luminosity 
\bee
L_{\nu_e}
=
4 \pi r^2 
\int \chi \omega f_{\nu_e} (\omega, \vec r) \frac{d^3k}{(2 \pi)^3}  .
\ene
Here $r$ is the distance from the center of the remnant to a given point.

Under the additional simplifying assumption 
\bee
\left< \omega_{\nu_e}^2 \right> L_{\nu_e} / \left< \chi_{\nu_e} \right> 
= 
\left< \omega_{\tilde\nu_e}^2 \right> L_{\tilde\nu_e} / 
  \left< \chi_{\tilde\nu_e} \right>  ,
\nonumber
\ene
which holds for various supernova explosion models, we obtain a well-known 
(see, e.g., Janka 2001) expression for the shock heating rate in the 
direct URCA processes: 
\bee
Q_0 
&=&
Q_0^{\nu} + Q_0^{\tilde\nu}
=
\frac { G^2  } {\pi}
(1 + 3 g_a^2)
\frac { L_{\nu_e} \left< \omega_{\nu_e}^2 \right> } 
      { 4 \pi r^2 \left< \chi_{\nu_e} \right> }
\simeq
\label{Qur}
\\
&\simeq&
55 \left( \frac{MeV} {s \cdot nucleon} \right) 
\left( \frac{L_{\nu_e}} { 10^{52} erg/s } \right) 
\left( \frac{ \left< \omega_{\nu_e}^2 \right> } 
       { 225 MeV^2 } \right) 
\left( \frac{ 10^7 cm } {r} \right)^2   ,
\nonumber
\ene
where $r$ is the characteristic distance to the shock.

The heating rate of the medium per nucleon in the additional processes 
(\ref{Nu1})-(\ref{Nu2}) can be calculated as  
\bee
Q_B^{\nu_i, \tilde\nu_i}
= 
\frac{1}{N_N} \int
E_{\nu_i, \tilde\nu_i} f_{\nu_i, \tilde\nu_i} (\omega, \vec r) 
\frac{d^3k}{(2 \pi)^3}  , 
\label{QB}
\ene
where $E_{\nu_i, \tilde\nu_i}$ is the heating rate of the medium per 
(anti)neutrino of type $i$. In the case of a moderate magnetic field where 
$ \left< \omega^2_{\nu_i} \right> \gg m_e^2 \gtrsim  eB $, 
it can be represented in 
a logarithmic approximation (Kuznetsov and Mikheev 1997) as 
\bee
E_{\nu_i, \tilde\nu_i}
\simeq
\frac{ 7 G_F^2 (c_{v_i}^2 + c_{a_i}^2) } {432 \pi^3} 
(eB \omega_{\nu_i} \sin\varphi)^2 
\ln \left( \frac{eB \omega_{\nu_i} \sin\varphi}{m_e^3} \right)  .
\ene
Here, $\varphi$ is the angle between the momentum of the initial neutrino 
and the magnetic field, 
$c_{v_i}$ и $c_{a_i}$ ($ c_{v_e} \simeq 0.96 $, $c_{a_e} = 1/2 $ 
for the electron  neutrino; 
$ c_{v_i} \simeq -0.04 $, $c_{a_i} = -1/2 $ for the
$\mu$- and $\tau$-neutrino).

However, the magnetic field $ B \gg B_0 = m_e^2 / e $ can be generated in 
a supernova shock wave. In the case of relatively strong magnetic field 
where  $ \left< \omega^2_{\nu_i} \right> \gg 2 eB \gg m_e^2 $, 
the heating rate was calculated as
\bee
E_{\nu_i, \tilde\nu_i}
\simeq
\frac{ 7 G_F^2 (c_{v_i}^2 + c_{a_i}^2) } {216 \pi^3} 
(eB \omega_{\nu_i} \sin\varphi)^2 
\ln \left( \frac{ \omega_{\nu_i}^2 \sin^2\varphi } {eB} \right)  .
\ene
This formula was obtained in a logarithmic approximation also. 
Using this expression, formula (\ref{QB}) for the neutrino heating rate of 
the medium in processes (\ref{Nu1})-(\ref{Nu2}) can also be expressed in 
terms of the mean quantities of neutrino radiation and its total 
luminosity:
\bee
Q^i_B 
=
Q_B^{\nu_i} + Q_B^{\tilde\nu_i}
=
\frac{ 7 G_F^2 (c_{v_i}^2 + c_{a_i}^2) } {108 \pi^3} 
\frac{ (eB)^2 \left< \omega_{\nu_i} \right> } {N_N}
\frac{L_{\nu_i}} { 4 \pi r^2 \left< \chi_{\nu_i} \right>}   
\ln \left( \frac{ \left< \omega_{\nu_i}^2 \right> } {eB} \right)  .
\ene
When deriving this formula, we assumed that $ \sin\varphi \sim 1 $, which 
is right approximately for the region where the neutrinos propagate almost freely. 
Under the additional simplifying assumption  
\bee
\left< \omega_{\nu_i} \right> L_{\nu_i} / \left< \chi_{\nu_i} \right> 
= 
\left< \omega_{\tilde\nu_i} \right> L_{\tilde\nu_i} / 
  \left< \chi_{\tilde\nu_i} \right> 
=
\left< \omega_{\nu_e} \right> L_{\nu_e} / \left< \chi_{\nu_e} \right>  ,
\nonumber
\ene
which holds good for various supernova explosion models, we obtained the 
following expression for the ratio of the total neutrino heating rates in 
processes (\ref{Nu1})-(\ref{Nu2}) and (\ref{Ur1})-(\ref{Ur2}):  
\bee
\frac{Q_B}{Q_0}
\simeq
1.0 \times 10^{-2} \;
\frac{ (eB)^2 \left< \omega_{\nu_e} \right> } 
     { N_N \left< \omega^2_{\nu_e} \right> } 
\simeq
\frac{ 9 MeV } { \left< \omega_{\nu_e} \right> } \;
\frac{(eB)^2}{\rho}  ,
\label{frac}
\ene
where $ Q_B = \sum\limits_i Q^i_B $ is the total neutrino heating rate for 
all types of neutrinos.

In this paper, we considered the most significant neutrino shock heating 
processes in the magnetorotational model. For the sake of generality, we 
derived the well-known expression for the heating rate in the URCA 
processes~(\ref{Qur}). Note that even this expression contains a number of 
simplifying assumptions discussed in the paper. In addition, the density 
of the medium decreases with distance much more slowly in the 
magnetorotational model than in the spherically symmetric model. This 
implies that, even at the characteristic distances where the shock is 
formed ($ r \sim 100 $~km), we must also take into account the neutrino 
radiation processes, which can significantly reduce the total rate of 
neutrino shock heating. Thus, formula (\ref{Qur}) should be treated with 
caution, particularly in the magnetorotational model.

On the other hand, in the magnetorotational model the strong 
magnetic field ($ B \sim 10^{15} $~G) can be generated at large distances 
($ r \sim 100 $~km). Therefore, we must consider the effect of such a 
strong magnetic field on the neutrino shock heating. In particular, under 
these conditions, the new neutrino heating reactions 
(\ref{Nu1})-(\ref{Nu2}) can compete with the direct URCA processes 
(\ref{Ur1})-(\ref{Ur2}), which are the main processes in the spherically 
symmetric explosion model.

Our estimate (\ref{frac}) shows that the new neutrino heating reactions 
(\ref{Nu1})-(\ref{Nu2}) become significant when $ (eB)^2 \gtrsim \rho $. 
However, the strength of the magnetic field produced by the medium cannot 
be too large. For example, in the models with sub-Keplerian rotation rates 
(Akiyama et al. 2003; Thompson et al. 2004), the magnetic field strength 
reaches saturation when the field energy density becomes comparable to the 
rotation energy density of the medium, 
$ B_{sat}^2 \simeq 4 \pi \rho (r \Omega)^2 $ (where $\Omega$ is the local 
angular velocity of the medium at distance $r$). Using this estimate, we 
can present the ratio of the heating rates (\ref{frac}) as  
\bee
\frac{Q_B}{Q_0}
\lesssim
0.1 \left( \frac{ r \Omega } {c} \right)^2 
\frac{ 9 MeV } { \left< \omega_{\nu_e} \right> }  \ll 1  ,
\ene
where $c$ is the speed of light in a vacuum. Thus, the new reactions that 
open up in a magnetic field cannot compete with the standard neutrino 
shock heating processes. Consequently, even in the case of a strong 
magnetic field, $ B \sim 10^{15} $~G, the heating is almost completely 
determined by the absorption and emission of neutrinos in the direct URCA 
processes, with the influence of the magnetic field on them being reduced 
to insignificant corrections.

ACKNOWLEDGMENTS 

We are grateful to N.V.~Mikheev, G.S.~Bisnovatyi-Kogan, and S.G.~Moiseenko 
for their discussions and advice on the fundamental points of this work. 
This work supported in part by the Council on Grants by the President 
of Russian Federation for the Support of Young Russian Scientists and 
Leading Scientific Schools of Russian Federation under the Grant 
No. NSh-6376.2006.2, and by the Russian Foundation for Basic Research 
under the Grant No. 04-02-16253.


\end{document}